\documentclass[12pt,a4paper]{article}

\usepackage{amsmath}

\newcommand{\eq}[1]{eq.(\ref{#1})}

\newcommand{\hepth}[1]{{\tt hep-th/#1}}
\newcommand{\plb}[3]{Phys.Lett. {\bf B#1} (#2) #3}
\newcommand{\npb}[3]{Nucl.Phys. {\bf B#1} (#2) #3}

\newcommand{\ep}{\varepsilon} % boundary parameter
\newcommand{\lam}{\lambda}
\renewcommand{\th}{\Theta}
\newcommand{\bl}[1]{\left(#1\right)}
\newcommand{\nn}{\nonumber}
\newcommand{\abs}[1]{\left|#1\right|}
\newcommand{\im}[1]{\text{Im}(#1)}

\advance\textheight by \topskip
\makeatletter

\@addtoreset{equation}{section}
\def\section{\@startsection {section}{1}{\z@}{-8.5ex plus -1ex minus
 -.2ex}{3.3ex plus .2ex}{\large\bf\centering}}
\def\subsection{\@startsection{subsection}{2}{\z@}{-3.25ex plus
 -1ex minus -.2ex}{1.5ex plus .2ex}{\bf}}
\def\subsubsection{\@startsection{subsubsection}{3}{\z@}{-3.25ex plus%
 -1ex minus -.2ex}{1.5ex plus .2ex}{\sl}}

\begin{document}

\begin{titlepage}
\vspace*{-2cm}

\begin{flushright}
KCL-MTH-99-27\\ DTP--99/43\\ hep-th/9909145 \\
\end{flushright}
\vspace{0.3cm}
\begin{center}
{\Large {\bf Boundary breathers
%\\ \vspace{2mm}
in the sinh-Gordon model }}\\ \vspace{1cm} {\large \bf E.\
Corrigan\footnote{\noindent E-mail: {\tt Edward.Corrigan@durham.ac.uk}}${}^*$
and G.\ W.\ Delius\footnote{\noindent E-mail: {\tt
delius@mth.kcl.ac.uk},~~~~home page: {\tt
http://www.mth.kcl.ac.uk/\~{}delius/}}${}^*$ }\\ \vspace{0.3cm} {${}^a$}\em
Department of Mathematical Sciences\\ Durham University\\ Durham DH1 3LE,
U.K.\\ \vspace{0.3cm} {${}^b$\em \it Department of Mathematics\\ King's
College London\\ Strand, London WC2R 2LS, U.K.}\\ \vspace{1cm} {\bf{ABSTRACT}}
\end{center}
\begin{quote}
We present an investigation of the boundary breather states of the sinh-Gordon
model restricted to a half-line. The classical boundary breathers are
presented for a two parameter family of integrable boundary conditions.
Restricting to the case of boundary conditions which preserve the
$\phi\rightarrow -\phi$
 symmetry of the bulk
theory, the energy spectrum of the boundary states is computed in two ways:
firstly, by using the bootstrap technique and subsequently, by using a WKB
approximation. Requiring that the two descriptions of the spectrum agree with
each other allows a determination of the relationship between the boundary
parameter, the bulk coupling constant, and the parameter appearing in the
reflection factor derived by Ghoshal to describe the scattering of the
sinh-Gordon particle from the boundary.
\end{quote}
\vspace{4cm}* Address after 1 October 1999:

{\em Department of Mathematics,
University of York, Heslington, York YO10 5DD, UK.}

\vfill

\end{titlepage}

\section{Introduction}
\label{s:intro}

In recent years, there has been renewed interest in field theories
defined on restricted domains. In particular, integrable
two-dimensional models, for example affine Toda field theories,
may be confined to a half-line or an interval by boundary
conditions which maintain integrability
\cite{Che,Skly88,Ghosh94a,Cor94a}
(for a partial review, see \cite{Cor96}). The variety of possibilities is
intriguing although in most Toda theories the freedom to choose boundary
conditions is severely limited to a finite, discrete set of possibilities.
In fact, within the models based on the $a_n^{(1)},\ d_n^{(1)}$
or $e_n^{(1)}$ data, only the model based on $a_1^{(1)}$,
the sinh-Gordon model,
allows parameters to be introduced as part of the boundary conditions
\cite{Bow96a}.

An outstanding question concerns the quantum integrability
of models with boundaries and although there has been some
progress towards understanding particular examples, mostly within
the class of models based on the $a_n$ series,
there remains much to be done
to discover the systematics underpinning the apparently
bewildering variety of cases.

Even within the sinh/sine-Gordon model, about which so much is now known,
there remain some open questions. Up to the present there appears to be a gap
in the understanding of how the boundary data, which is prescribed in order to
formulate the boundary conditions of the model, is related to the parameters
appearing in the family of reflection factors describing particle-boundary
scattering. Finding this relationship needs the answers to dynamical questions
which cannot be resolved by general requirements such as the reflection
Yang-Baxter equation, or `crossing-unitarity'. In this article we shall
examine the sinh-Gordon model restricted to a half-line by boundary conditions
preserving its bulk symmetry and for which one expects boundary bound states,
and we shall approach the boundary bound states from two points of view. On
the one hand, we will calculate their spectrum using a semi-classical approach
rooted in the classic work of Dashen, Hasslacher and Neveu \cite{DHN75} while,
on the other, we will compute the same data using bootstrap techniques. The
marriage of the two approaches will yield strong evidence for a conjectured
relationship between the reflection factors and the boundary data.

\section{The sinh-Gordon model on the half line}
\label{s:sg}

The sinh-Gordon model describes a single real scalar field $\phi$
in 1+1 dimension with exponential self-interaction. The field
equation is
\begin{equation}\label{fe}
  \partial_t^2\phi-\partial_x^2\phi+
  \frac{\sqrt{8}m^2}{\beta}\sinh(\sqrt{2}\beta\phi)=0,
\end{equation}
where $m$ and $\beta$ are parameters and we have used normalisations
customary in affine Toda field
theories of which the sinh-Gordon model is the simplest example
\cite{Bra90}.
The dimensional mass parameter $m$ will be set to unity.

In contrast to the sine-Gordon model with its soliton and breather
solutions the sinh-Gordon model has only one real non-singular
classical
solution, namely the constant vacuum solution $\phi=0$. In the
quantum theory the small oscillations around this vacuum
correspond to the sinh-Gordon particle.

The sinh-Gordon model is integrable which implies in particular
that there are infinitely many independent conserved charges $Q_{\pm s}$,
 where $s$ is
any odd integer, and the S-matrix describing the scattering of $n$
sinh-Gordon
particles factorises into a product of $n(n-1)/2$ two-particle
scattering amplitudes. The scattering  between two
particles of relative rapidity $\th$ is conjectured to be given by
the S-matrix factor \cite{Fadd78, Zam79}
\begin{equation}\label{s}
  S(\th)=-\frac{1}{(B)(2-B)},
\end{equation}
where we have used the convenient block notation \cite{Bra90}
\begin{equation}\label{blo}
  \bl{x}=\frac{\sinh\left(\frac{\th}{2}+\frac{i\pi x}{4}\right)}
  {\sinh\left(\frac{\th}{2}-\frac{i\pi x}{4}\right)},
\end{equation}
and the coupling constant $B$ is related to the bare coupling constant
$\beta$ by $B=2\beta^2/(4\pi+\beta^2)$. Traditionally, scattering
and other properties
of the sinh-Gordon model have been obtained from knowledge of the
lowest breather in the sine-Gordon model
  by analytic continuation in the coupling constant (but, see also
\cite{Skly89}).

The sinh-Gordon model can be restricted to the left half-line
$-\infty\leq x\leq 0$ without losing integrability by imposing
the boundary condition
\begin{equation}\label{bc}
  \left. \partial_x\phi\right|_0=\frac{\sqrt{2}m}{\beta}
  \left(\ep_0e^{-\frac{\beta}{\sqrt{2}}\phi(0,t)}-
\ep_1e^{\frac{\beta}{\sqrt{2}}\phi(0,t)}\right) ,
\end{equation}
where $\ep_0$ and $\ep_1$ are two additional parameters \cite{Ghosh94a,
MacI95}. This
set of boundary conditions generally breaks the reflection symmetry
$\phi\rightarrow -\phi$ of the sinh-Gordon model. However, the  symmetry is
preserved when $\ep_0=\ep_1\equiv\ep$ and much of this article will be
devoted to this special case.

To describe the sinh-Gordon particles on the half line one needs in addition
to the two-particle scattering amplitude \eqref{s} also the amplitude for the
reflection of a single particle from the boundary. This reflection amplitude
can be deduced from the lowest breather reflection amplitude in the
sine-Gordon model by analytic continuation in the coupling constant (i.e., the
continuation $\lam\rightarrow -2/B$ in the notation of \cite{Ghosh94a}). Using
the breather reflection amplitudes calculated by Ghoshal \cite{Ghosh94b}
gives\footnote{In Ghoshal's notation $E=B\eta/\pi, \ F=iB\vartheta/\pi$.}
\begin{equation}\label{br}
  K(\theta,\ep_0,\ep_1, \beta)=\frac{(1)(2-B/2)(1+B/2)}
{(1-E(\ep_0,\ep_1,\beta ))(1+E(\ep_0,\ep_1,\beta ))
(1-F(\ep_0,\ep_1,\beta ))(1+F(\ep_0,\ep_1,\beta ))},
\end{equation}
where we are again using the block notation from \eqref{blo} but
in \eqref{br} $\theta$
represents the rapidity of a single particle. When the bulk
reflection symmetry is preserved one of the two parameters $E$ or $F$
vanishes. We shall choose $F=0$, and consequently  one obtains
\begin{equation}\label{brsymm} K_0(\theta,\beta )\equiv K(\theta,\ep,
\beta)=\frac{(2-B/2)(1+B/2)}{(1) (1-E(\ep,\beta ))
(1+E(\ep,\beta ))}\equiv
K_D \frac{1}{(1-E)(1+E)}.
\end{equation} Actually, the first factor, $K_D$
is the reflection factor corresponding to the Dirichlet boundary condition
$\phi (0,t)=0$, as noted in \cite{Ghosh94a,Ghosh94b}. All reflection factors
satisfy the crossing-unitarity relation which, in the case of
scalar reflection factors, reads,
\begin{equation}\label{cu}
K\left(\theta + \frac{i\pi}{2}\right) K\left(\theta-\frac{i\pi}{2}
\right) S(2\theta)=1.
\end{equation}
The Dirichlet reflection factor $K_D$ satisfies \eqref{cu} by itself.

In this paper we note that contrary to the situation on the whole line, the
sinh-Gordon equation restricted to a half-line by integrable boundary
conditions has non-singular, finite energy, breather solutions. After
quantising, these will lead to a spectrum of boundary bound states which ought
to match not only the physical strip poles of the expression \eqref{brsymm}
but also the poles appearing in similar expressions derived from
\eqref{brsymm} using the boundary bootstrap. These derived reflection factors
will be determined below and represent
 the sinh-Gordon particle
reflecting from the excited  boundary states. Matching the  two ways of
looking at the energies of the excited states will determine a relation
between $\ep, \ \beta$ and $E$, see \eq{E}. In fact the relationship between
the two parameters coincides with a tentative suggestion made in \cite{Cor97}.

A similar analysis is feasible in the general case ($\ep_0\ne\ep_1$)
but it will not be carried out here. However, the associated boundary
breathers and a few of their properties will be described as an essential
preliminary to a fuller investigation.

\section{Boundary Breathers}
\label{s:bb}

The sinh-Gordon model on the whole line has no
non-singular real solutions other than $\phi =0$. However,
there are singular real
breather solutions satisfying the boundary condition \eqref{bc}
whose singularities can be designed to lie for all time
on the right half
line ($x>0$).  Thus, they are well-defined periodic solutions of the
sinh-Gordon model on the left half-line and we shall call them  {\it
boundary breathers}. Following Hirota, with suitable choices of $\tau_j$,
the solutions can be written conveniently in the form \cite{Hiro80}
\begin{equation}\label{st}
  \phi=\frac{\sqrt{2}}{\beta}\ln\frac{\tau_0}{\tau_1}.
\end{equation}
 For  the symmetrical boundary conditions with $\ep_0=\ep_1=\ep$,
appropriate choices are:
\begin{equation}\label{bt}
  \tau_j=1+(-1)^j 2\cos(2t\sin\rho )e^{2x\cos\rho }\frac{1}{\tan\rho}
  \sqrt{\frac{\ep+\cos\rho}{\ep-\cos\rho}}-
  e^{4x\cos\rho }\left(\frac{\ep+\cos\rho}{\ep-\cos\rho}\right),
\end{equation}
where the parameter $\rho$ determines the frequency of the breather.

 In order for the  $\tau$ functions to be real,
the square root appearing in \eqref{bt} must be,
which in turn requires that
$\abs{\ep}\geq \cos\rho$.
Also, the solution will be  singular whenever one of
the $\tau$ functions is
zero. While singularities cannot be avoided entirely it is possible
to ensure that there are none in the region $x\le0$, and that is
sufficient
for the present purpose. Noting that for a particular $x$
singularities cannot  occur at any time
provided
\begin{equation}\label{}
1<\left|\frac{\tan\rho\left(1- e^{4\cos\rho x}\left(\frac{\ep+\cos\rho}
{\ep-\cos\rho}\right)\right)}{ 2 e^{2\cos\rho x}\sqrt{\frac{\ep+
\cos\rho}{\ep-\cos\rho}}}\right|,
\end{equation}
we deduce that requiring there are no singularities on the left half-line
is equivalent to the restrictions
\begin{equation}\label{co}
  -1<\ep<0~~~\text{ and }~~~\cos\rho<-\ep.
\end{equation}
Note, at $\cos\rho=-\ep$ the solution collapses to the vacuum solution
$\phi=0$ indicating that there is a minimum allowed frequency for
a breather which
is strictly greater than zero. This is  a distinctive feature not
shared by the usual breathers of the sine-Gordon model on the full
line whose frequencies may approach zero.

The energy functional of the sinh-Gordon model with boundary
condition \eqref{bc} is
\begin{align}\label{en}
  \cal{E}[\phi]=&\int_{-\infty}^0 dx\left(\frac 12\dot{\phi}^2+
  \frac 12{\phi'}^2+\frac{2}{\beta^2}
  \left(\cosh(\sqrt{2}\beta\phi)-1\right)\right)\nn\\&~~+
  \frac{2}{\beta^2}\left(\ep_0 (e^{-\frac{\beta}{\sqrt{2}}\phi(0,t)}-1)
+\ep_1 (e^{\frac{\beta}{\sqrt{2}}\phi(0,t)}-1)\right),
\end{align}
but it is most easily calculated in terms of the $\tau$ functions
as a boundary term \cite{Del98a},
\begin{equation}\label{enb}
  {\cal{E}}[\phi]=\frac{2}{\beta^2}\left.\left(\ep_0
\left(\frac{\tau_0}{\tau_1}-1\right)
+\ep_1\left(\frac{\tau_1}{\tau_0}-1\right)-\left(\frac{\tau_0^\prime}{\tau_0}
+\frac{\tau_1^\prime}{\tau_1}\right)\right)\right|_{\rm x=0}.
\end{equation}
The energy of the real boundary
breather turns out to be given by
\begin{equation}\label{eb}
  {\cal E}_{\text{breather}}=\frac{8}{\beta^2}(-\cos\rho-\ep),
\end{equation}
and the condition \eqref{co} ensures $ {\cal E}_{\text{breather}}$
is always positive, or zero if $\cos\rho =-\ep$.

In the quantum theory, the continuum of boundary breather
solutions is expected to lead to a discrete spectrum of boundary
states. To obtain an estimate for the energies of these
boundary states one might in the first instance use
 the Bohr-Sommerfeld quantisation condition (see for example
\cite{Col}). One proceeds directly by calculating the left
hand side of
\begin{equation}\label{BS}
\int_0^Tdt \int_{-\infty}^0 dx\, \pi (x,t) \,
\dot{\phi} (x,t)=(2n+1)\pi,
\end{equation}
where $\pi (x,t)=\dot{\phi}$ is the momentum conjugate to $\phi$,
$T=\pi/\sin\rho$
is the period of the breather, and $n$ is an integer. (As usual,
we have put $\hbar=1$).

It is convenient to set
$\ep =\cos \pi a, \  1>a>1/2$, implying that
$\cos^{-1}(-\ep)= \pi(1-a)$, then integrating gives
\begin{equation}\label{KE}
\int_0^Tdt \int_{-\infty}^0 dx\, \dot{\phi}^2= \frac{8\pi}{ \beta^2}
(\rho -\pi (1-a)).
\end{equation}
The energy levels follow from \eqref{BS} yielding
 \begin{equation}\label{spectrum}
 {\cal E}_n=\frac{8}{ \beta^2}\left(-\cos\pi a
 +\cos{\pi}\left((n+1/2)\frac{\beta^2}{ 4\pi} -a\right)
\right).
\end{equation}

Since the breathers approach the vacuum solution as $\rho\rightarrow
\pi(1-a)$ by reducing their amplitudes to zero rather than their
frequencies,
it is natural that the boundary breather spectrum should have a
zero point energy. This is the reason for the postulated form
of the right hand side of  \eqref{BS}.

The difference between two consecutive bound state energies is readily
deduced from \eqref{spectrum} and conveniently written,
\begin{equation}\label{ediffs}
 {\cal E}_{n+1}-{\cal E}_n=
\frac{16}{ \beta^2}\sin\left(\frac{\beta^2}{
8}\right) \cos\frac{\pi}{
 2}\left(\left(n+1\right)\left(\frac{\beta^2}{ 2\pi}\right)-
 2a+1\right).
\end{equation}

It will be seen below that this has the form we
would have anticipated from the boundary bootstrap. However, given that we
know the coupling constant renormalises, and we expect the boundary coupling
to renormalise too \cite{Pen96}, the outcome of this calculation
can be at best an indication. A more reliable method for quantising the
boundary breathers is likely to be an adaptation of the techniques developed
by Dashen, Hasslacher and Neveu \cite{DHN75} and this will be pursued in
section 5.

Finally, we shall end this section with a brief description of the boundary
breathers in those cases where the boundary conditions break the
bulk symmetry.
As before, the
solutions have the general form indicated by \eqref{st} but this time
the two tau
functions are more elaborate and given by
\begin{eqnarray}\label{stasymm}
\tau_j&=&1+(-1)^j\left(2\frac{s}{\tan\rho}\cos(2t\sin\rho )\exp{(2x\cos\rho )}
 + \right. \nonumber\\
& &\ \ \ \left.\frac{r}{\tan^2\frac{\rho}{ 2}}\, \exp(2x) -r
s^2 \, \tan^2\frac{\rho}{ 2}\exp\, 2x(2\cos\rho +1) \right)\\
& &\ \ \ \ \ \ \  -\left(2\frac{r\, s}{\tan\rho}\cos(2t\sin\rho )\exp\,
2x(\cos\rho +1)
 +  s^2\exp (4x\cos\rho)\right),\nonumber
\end{eqnarray}
where,
\begin{eqnarray}
r&=&\frac{\sin\frac{\pi a_0}{2}-\sin\frac{\pi a_1}{2}
}{ \sin\frac{\pi a_0}{2}+\sin\frac{\pi a_1}{2}}\nonumber\\
& & \\
s^2&=&\frac{\left(1+\cos\rho\right)\left(\cos\frac{\pi(a_0+a_1)}{2}
+\cos\rho\right)\left(\cos\frac{\pi(a_0-a_1)}{2}
-\cos\rho\right)}{ \left(1-\cos\rho\right)\left(\cos\frac{\pi(a_0+a_1)}{2}
-\cos\rho\right)\left(\cos\frac{\pi(a_0-a_1)}{2}
+\cos\rho\right)}\nonumber
\end{eqnarray}
and $a_0$ and $a_1$ are related to the boundary parameters by,
\begin{equation}
\ep_0=\cos\pi a_0, \quad \ep_1=\cos\pi a_1.\nonumber
\end{equation}
Numerical investigation of these boundary breathers indicates that they
are non-singular in the region $x<0$ provided
\begin{equation}
0<\cos\rho <-\cos\frac{\pi (a_0+a_1)}{2}, \qquad \cos\frac{\pi (a_0+a_1)}{2}<0,
\qquad \cos\frac{\pi (a_0-a_1)}{2}>0;\nonumber
\end{equation}
their energies are given by
\begin{equation}
{\cal E}=\frac{4}{\beta^2}\left(-2-2\cos\rho + \left(\sin\frac{\pi a_0}{2}
+\sin\frac{\pi a_1}{2}\right)^2\right).
\end{equation}
Again, the breathers have frequencies bounded below because the $(a_0,
a_1)$ parameters are restricted. For example, they could lie within the ranges
$ -1<a_0-a_1<1, \ 1<a_0+a_1<2$ in the positive quadrant. The
boundary breathers for boundary conditions preserving the symmetry of
the sinh-Gordon equation are included as the special case $a_0=a_1$. The
possibility $a_0=-a_1$ is outside the range.

These solutions may be considered as a superposition of a static `soliton'
and a `boundary breather', carefully designed to be real and non-singular,
and
to satisfy the general boundary condition \eqref{bc}. They are sinh-Gordon
counterparts of the sine-Gordon solutions considered by Saleur, Skorik and
Warner \cite{Sal94}.

\section{The boundary bootstrap}
\label{s:boot}

For certain ranges of values of the parameters $E$ and $F$ the particle
reflection amplitude \eqref{br} has simple poles at particular values of
$\theta$ on the physical strip, $0<\im{\theta}<\pi/2$. For the case $F=0$
these must\footnote{ Because there is no three-point coupling in the
sinh-Gordon model with symmetrical boundary condition, simple poles on the
physical strip can never be due to a generalised Coleman-Thun mechanism
\cite{dorey98}.} be due to the propagation of virtual excited boundary states.
The amplitudes for the reflection of the sinh-Gordon particle from these
excited boundary states is obtained by the boundary bootstrap \cite{Cor94a,
Ghosh94a, Fring95}. When the reflection factor \eqref{br} has a pole at
$\theta=i\psi$ with $0<\psi<\pi /2$ then the reflection factor corresponding
to the associated excited boundary state is calculated via the relation
\begin{equation}\label{bbs}
K_1 (\theta )= K_0(\theta ) S(\theta - i\psi )S(\theta + i\psi),
\end{equation}
where $S(\theta)$ is the two-particle S-matrix \eqref{s}.
Also, since energy is conserved, the energy of the excited boundary state is
given by \begin{equation}\label{excite} {\cal E}_1={\cal E}_0 +
m(\beta)\cos\psi , \end{equation} where $m(\beta )$ is the mass of the
sinh-Gordon particle.

Considering the case $F=0$, the  regions in E  where the amplitude
\eqref{brsymm} has poles on the physical strip are
\begin{equation}
{\rm I}: 2>E>1 \quad \hbox{and} \quad {\rm II}: -2<E<-1;
\end{equation}
since  $0\le B\le 2$, the other factors (in $K_D$) never have poles in
the physical strip. In region I,  $\psi=\pi (E-1)/2$ and, using \eqref{bbs},
we derive the reflection factor for the first excited state,
\begin{equation}
K_1=K_D\, \frac{1}{(1-E)(1+E)}\, \frac{(1+E+B)(1-E-B)}{(1-E+B)(1+E-B)}.
\end{equation}
This in turn has a new pole at $\psi=\pi (E-1-B)/2$ provided $B<E-1$
indicating another excited state whose reflection factor is
\begin{equation}
K_2=K_D\, \frac{1}{(1-E+B)(1+E-B)}\, \frac{(1+E+B)(1-E-B)}{(1-E+2B)(1+E-2B)}.
\end{equation}
Continuing in this vein leads to a set of excited states with
associated reflection factors given by,
\begin{equation}\label{Kn}
K_n=K_D\, \frac{1}{(1-E+(n-1)B)(1+E-(n-1)B)}\, \frac{(1+E+B)(1-E-B)}{(1-E+nB)
(1+E-nB)}.
\end{equation}
Note that the pole corresponding to the ($n+1$)st state will be within
the correct range provided $E$ satisfies $ 2>E>1+nB $. Thus, for a given
$E$ and $B$ there can be at most a finite number of bound states, and
possibly none. Note too that the reflection factor for scattering
from the $n$th bound state also contains a pole corresponding
to the ($n-1$)st bound state. We shall see that there is a subtlety
concerning the coefficient of this pole because it develops a zero
at an $n$-dependent critical value of $\beta$.

The energies of the boundary states are given by repeatedly applying
\eqref{excite} and satisfy,
\begin{equation}\label{ediff}
{\cal E}_{n+1}={\cal E}_n + m(\beta )\cos \frac{\pi}{2} (nB-E+1).
\end{equation}
This is the result that we wish to compare with the quantisation of the
classical breather spectrum in order to determine
$E(\ep,\beta)$ and $m(\beta)$. However, we shall defer
the comparison until after we have developed the Dashen,
Hasslacher, Neveu argument in the present context.

\section{Semi-classical quantisation}
\label{s: semi}

To carry out the semi-classical calculation it is first necessary
to solve the sinh-Gordon equation linearised in the presence of the
boundary breathers. Setting $\phi =\phi_0+\eta$, the linear wave
equations for the fluctuations are:
\begin{equation}\label{linear}
\frac{\partial^2\eta}{\partial t^2}-\frac{\partial^2\eta}{\partial x^2}
+4 \eta \cosh\sqrt{2}\beta\phi_0 =0, \qquad
\left(\frac{\partial \eta}{
\partial x}+ 2\ep \eta\cosh \frac{\beta\phi_0}{ \sqrt{2}}\right)_{\rm
x=0}=0.
\end{equation}
It is convenient to solve \eqref{linear} by
perturbing \eqref{st}; in other words, we take
\begin{equation}
\eta=\frac{\tau_1\delta\tau_0-\tau_0\delta\tau_1}{ \tau_0\tau_1},
\end{equation}
with $\delta\tau_j$
chosen as follows:
\begin{eqnarray}\label{deform}
\tau_j+\delta\tau_j&=&1+(-)^j\left(( e_1 +  e_2 + E_1 +
E_2\right) +A_{12} E_1 E_2 +e_1(\mu_{11}E_1+\mu_{12}E_2)\nonumber\\
& & \\
& & \ \ \ \  +e_2(\mu_{21}E_1
+\mu_{22}E_2)+(-)^j A_{12}E_{12}(\mu_{11}\mu_{12} e_1 +\mu_{21}\mu_{22}
e_2).\nonumber
\end{eqnarray}
In \eqref{deform} we have made use of the Hirota expression for the
general multi-soliton solution to the sine-Gordon equation \cite{Hiro80,
Holl92}, suitably adapted to solve the sinh-Gordon equation,
keeping terms up
to first order in $e_1$ and $e_2$. The various quantities
are given by:
\begin{eqnarray}
e_1&=&\lambda_1\, e^{-i\omega t +ikx}, \ \ e_2=\lambda_2\,
e^{-i\omega t -ikx},\ \ \ \ \ \omega^2-k^2=4\nonumber\\
&&\nonumber \\
E_1&=&
\exp(2x\cos\rho +2it \sin\rho +x_0),\ \  E_2=
\exp(2x\cos\rho -2it \sin\rho +x_0)\nonumber\\
& & \nonumber \\
A_{12}&=&-\tan^2\rho, \ \ \ \ \ e^{x_0}=
\frac{1}{\tan\rho}\sqrt\frac{\ep+\cos\rho
}{\ep -\cos\rho}, \nonumber\\
& & \nonumber \\
\mu_{11}&=&\frac{1}{ \mu_{22}}=\frac{4+2\omega\sin\rho -2ik\cos\rho
}{ -4+2\omega\sin\rho -2ik\cos\rho},\ \ \mu_{12}=\frac{1}{
\mu_{21}}=\frac{-4+2\omega\sin\rho +2ik\cos\rho }{ 4+
2\omega\sin\rho +2ik\cos\rho},
\end{eqnarray}
where $\lambda_1$ and $\lambda_2$ are small parameters. Matching the
boundary condition at $x=0$ fixes the ratio $\lambda_2/\lambda_1$ to be
\begin{equation}\label{lratio}
K_B=\frac{\lambda_2}{\lambda_1}=\mu_{11}\, \mu_{12}\, \frac{ik+2\ep}{
ik-2\ep}=\frac{(ik+2\cos\rho)^2}{ (ik-2\cos\rho)^2}\, \frac{(ik+2\ep)}{
(ik-2\ep)}.
\end{equation} In the limit $x\rightarrow -\infty$,
\begin{equation}\label{breflect}
\eta\sim \lambda_1\, e^{-i\omega t}\left(e^{ikx}+K_B\, e^{-ikx}\right) ,
\end{equation}
defining the classical reflection factor corresponding to the boundary
breather. Taking $\cos\rho = -\ep$, the breather collapses to the
vacuum solution $\phi_0=0$ and the reflection factor collapses to
\begin{equation}\label{vreflect}
K_0=\frac{ik+2\ep}{ ik-2\ep}\equiv -\, \frac{1}{ (1-2a)(1+2a)}
\end{equation}
The ground state reflection factor is easily checked directly and it is the
$\beta\rightarrow 0$ limit of the reflection factor given in \eqref{brsymm}.
Hence, we may deduce that $E(0)=2a$.

The classical action of the boundary breather is calculated to be
\begin{equation}\label{claction}
S_{\rm cl}=\int_0^Tdt\int_{-\infty}^0dx\, {\cal L}=
\frac{8\pi}{\beta^2}\left(\rho- \pi (1-a)  +\frac{\cos\rho +
\cos\pi a }{ \sin\rho}\right),
\end{equation}
and vanishes as it should when $\cos\rho=-\ep$, i.e. $\rho =\pi (1-a)$.

The period $T=\pi /\sin\rho$ of the boundary breather defines the
`stability angles' via
\begin{equation}\label{stab}
\eta (t+T,x)=e^{-i\nu }\eta(t,x)\equiv e^{-i\omega T}\eta(t,x)
\end{equation}
and the field theoretical version of the WKB approximation makes
use of the stability angles together with a regulator to calculate
a quantum action. The standard procedure would be to place the field theory
in an interval $[-L,L]$ with periodic boundary conditions and to manipulate the
sum over the discrete stability angles so obtained. However, that option is not
available in this case. Instead, it is convenient to treat the sinh-Gordon
model in the interval $[-L,0]$ and to impose the Dirichlet condition
$\eta(t,-L)=0$. Since the limit $L\rightarrow\infty$ will  be taken
eventually,
the stability angles for the boundary breather ($\nu_B$), or the vacuum
solution ($\nu_0$) are effectively determined by the reflection factors
given in \eqref{breflect} or \eqref{vreflect}, respectively.

Following \cite{DHN75, Raj82} we need to calculate a sum over the stability
angles and use it to correct the classical action. Thus,
\begin{equation}\label{D}
\Delta=\frac{1}{ 2}\sum \, (\nu_B-\nu_0)\equiv \frac{T}{ 2}\sum\,
\left(\sqrt{ k_B^2+4}-\sqrt{k_0^2+4}\right), \end{equation} where $k_B$ and
$k_0$ are the sets of (discrete) solutions to
\begin{equation}\label{qconditions}
 e^{2ik_BL}=-\,
\frac{(ik_B+2\cos\rho)^2}{ (ik_B-2\cos\rho)^2}\, \frac{(ik_B+2\ep)}{
(ik_B-2\ep)},\ \ \ e^{2ik_0L}=-\, \frac{ik_0-2\ep}{ ik_0+2\ep}\, .
\end{equation}
Once $\Delta$ is known the quantum action is defined by
\begin{equation}\label{quaction}
S_{\rm qu}=S_{\rm cl}-\Delta.
\end{equation}

One way to proceed is to note that for large $k$ the solutions to
either of \eqref{qconditions} are close to

$$k_n=\left(n+\frac{1}{ 2}\right)\frac{\pi}{ L},$$
and so it is reasonable to set $(k_B)_n=(k_0)_n+\kappa((k_0)_n) /L$
where, for $L$ large, the function $\kappa$ is given approximately by
\begin{equation}\label{kappadef}
e^{2i\kappa (k)}=\frac{(ik+2\cos\rho)^2}{ (ik-2\cos\rho)^2}\,
\frac{(ik+2\ep)^2}{ (ik-2\ep)^2}\, .
\end{equation}
In terms of $\kappa$ the expression \eqref{D} is rewritten
\begin{equation}
\Delta\sim \frac{T}{ 2L}\sum_{n\ge 0}\, \frac{(k_0)_n\kappa ((k_0)_n)}{
\sqrt{ (k_0)_n^2+4}}+O(1/L^2),\nonumber \end{equation} and this, in turn, as
$L\rightarrow\infty$ can be converted to a convenient (but actually
divergent) integral, \begin{equation}\label{Da} \Delta =\frac{T}{
2\pi}\int_0^\infty\, dk \frac{k \kappa(k)}{ \sqrt{k^2+4}} \end{equation}
with which we shall have to deal.  Note that $\kappa$ vanishes when $\cos\rho
= -\ep$.

Integrating \eqref{Da} by parts we find
\begin{equation}\label{Db}
\Delta=\frac{T}{ 2\pi}\left( \left. \kappa \sqrt{k^2+4}\,
\right|^\infty_0
-\int_0^\infty dk \frac{d\kappa}{ d k} \sqrt{k^2+4}\right),
\end{equation}
where
\begin{equation}\label{kderv}
\frac{d\kappa}{dk} = \frac{4\cos\rho}{ k^2+4\cos^2\rho} +
\frac{4\cos\pi a}{
k^2+4\cos^2\pi a},
\end{equation}
and we note that with a suitable choice of branch
\begin{equation}\label{klimit}
\kappa\sim -\frac{4\cos\rho}{ k}-\frac{4\cos\pi a}{ k}\ \ \hbox{as}\ \
k\rightarrow\infty .
\end{equation}
From \eqref{klimit} and recalling that $\cos\rho < -\ep$, we deduce that
$\kappa$ approaches zero from above as $k\rightarrow
\infty$. Also, from \eqref{kderv} it is clear that the derivative of
$\kappa$ is positive  near $k=0$ but negative as $k\rightarrow \infty$.
Hence, the first term in \eqref{Db} is well-defined and the appropriate
branch of $\kappa$ has $\kappa (0)=0$. On the other
hand, the derivative of $\kappa$ is not decaying sufficiently rapidly
to ensure the
second term in \eqref{Db} is finite. However, this was to be expected
since a perturbative analysis of the sinh-Gordon model confined to
a half-line needs mass and
boundary counter terms to remove logarithmic divergences (which would
be removed automatically by normal-ordering the products of fields
in the bulk theory). With this in mind, the integral remaining in
\eqref{Db} should be replaced by \begin{equation}\label{counter}
\int_0^\infty dk\sqrt{k^2+4}\left(\frac{4\cos\rho}{ k^2+4\cos^2\rho}-
\frac{4\cos\rho}{ k^2+4}+\frac{4\cos\pi a}{ k^2+4\cos^2\pi a}-
\frac{4\cos\pi a}{ k^2+4}\right),
\end{equation}
the first counter-term removing the bulk divergence and the second
being there to remove a similar divergence associated with the boundary.
In effect, we are regarding the parameter $a$ as describing the bare
coupling which appears
in the boundary part of the Lagrangian once it is written in
terms of normal-ordered products of fields.
The counter-terms clearly respect the symmetry and the whole expression
vanishes when $\rho = \pi(1- a)$. The integrals in \eqref{counter} need
to be treated carefully with an eye to the facts that $\cos\rho >0$
but $\cos \pi a <0$.

Assembling the various components leads to
\begin{equation}\label{Dc}
\Delta= -\frac{2}{ \sin\rho}\left(\cos\rho +\cos\pi a +\rho \sin\rho
-\pi(1-a)\sin\pi a \right).
\end{equation}
 Recalling \eqref{claction}, and using \eqref{Dc}, the
quantum action defined in \eqref{quaction} is given by an expression of
the form
\begin{equation}\label{qactiona}
S_{\rm qu}=\frac{4}{ B}\left(\frac{\cos\rho}{\sin\rho} +\rho -\frac{\pi}{ 2}
\right) +\frac{8\pi}{\beta^2}\left(\pi a -\frac{\pi}{ 2}\right)
+\frac{\Gamma(a)}{ \sin\rho} +\pi, \end{equation} where $\Gamma$ is
independent of $\rho$,
\begin{equation}
\Gamma=\frac{4}{B}\cos\pi a +2\pi (a-1) \sin \pi a.
\end{equation}

Once the quantum action is determined the quantum energy is defined
by
\begin{eqnarray}\label{qenergy}
{\cal E}_{\rm qu}&=& -\frac{\partial S_{\rm qu}}{ \partial
T}=\frac{\sin^2\rho}{ \pi \cos\rho}\frac{\partial S_{\rm qu}}{\partial\rho}
=-\frac{4}{ \pi B}\cos\rho -\frac{\Gamma}{ \pi},
\end{eqnarray}
and the WKB
quantisation condition states that \begin{equation}\label{WKB} W_{\rm qu}=
S_{\rm qu}+T{\cal E}_{\rm qu}= \frac{4}{ B} \left(\rho -\frac{\pi }{
2}\right) +\frac{8\pi}{\beta^2}\left(\pi a- \frac{\pi}{ 2}\right)
+\pi=2N\pi .
\end{equation}
Here, $N=n+N_0$ with $n$ a positive integer or
zero, and we expect $N_0$ should be $1/2$.
Hence, the energies of the quantised boundary breather states are
 determined by a set of special angles $\rho_n$,
\begin{equation}
\rho_n=\frac{\pi}{2}\left(1+B\left(N-\frac{1}{2}\right) -
\frac{2\pi B}{\beta^2} (2a -1)\right),
\end{equation}
and
given by
\begin{equation}\label{qspectrum} {\cal E}_n=-\frac{4}{ \pi B}\cos
\rho_n - \frac{\Gamma}{ \pi} = -\frac{4}{ \pi B} \cos \frac{\pi}{
2}\left(\left(N-\frac{1}{ 2}\right)B + 1 -\frac{2\pi B}{ \beta^2}
(2a-1)\right) -\frac{\Gamma}{ \pi}.
\end{equation}
Notice that as
$\beta\rightarrow 0$, $\rho_n\rightarrow \pi (1-a)$ independently of $N$.
Thus, the frequencies collapse to the lowest allowed frequency, namely
$\omega_0=2\sin a\pi$.
On the other hand, in the same limit the energies are independent of $\beta$
and non-zero,
\begin{equation}\label{limitspectrum}
{\cal E}_n\rightarrow
N\omega_0.
\end{equation}
This is precisely the spectrum of a harmonic
oscillator vibrating at the fundamental frequency $\omega_0$
provided we set
$N=n+1/2$. With this interpretation, the
vacuum has a non-zero zero-point energy due to the presence of the
boundary.

Using \eqref{qspectrum} the corresponding differences in the
energy levels
are given by
\begin{equation}\label{qediff}
{\cal E}_{n+1}={\cal E}_{n} +\frac{8}{ \pi B}\sin \frac{\pi B}{ 4}\,
\cos \frac{\pi}{ 2}\left(\frac{2\pi B}{\beta^2}(2a-1)-NB\right).
\end{equation}
Comparing \eqref{qediff} with the outcome of the bootstrap calculation
\eqref{ediff} ought to assist us in identifying the unknown parameter
$E$ which appeared
in the expression for the reflection factor \eqref{brsymm}. Thus,
from the first excited level we deduce,
\begin{equation}\label{compare}
E-1=  \frac{2\pi B}{\beta^2}(2a-1)-N_0 B\equiv (2a-1) \left(1-\frac{B}{
2}\right)-N_0 B.
\end{equation}
Rearranging, we have
\begin{equation}\label{E}
 E(\ep,\beta)= 2a\left(1-\frac{B}{ 2}\right) + (1-2 N_0)
\frac{B}{ 2}.
\end{equation}

Taking the limit as $a\rightarrow 1/2$ from above, \eqref{E}
is in agreement
with the expression given by Ghoshal and Zamolodchikov for the Neumann
condition provided $N_0=1/2$ \cite{Ghosh94a}.
With $a$ arbitrary, \eqref{E}
agrees both with  perturbative calculations to order $\beta^2$ given in
\cite{Cor97, Top97}, and with a  conjectured all-orders guess reported in
\cite{Cor97}.  Once $N_0$ is chosen, the other excited states match up in the
two calculations without any further adjustments.

In the bulk sine-Gordon theory the analogous quantity to $N_0$ vanishes
in the Dashen, Hasslacher, Neveu calculation of the breather spectrum.
In the half-line theory, we have found that the two ways of regarding the
spectrum of boundary bound states match provided $N_0=1/2$. Although we
do not yet have an independent reason for expecting  this value of
$N_0$ on the basis of WKB theory, its appearance in \eqref{WKB} is reminiscent
of the extra $1/2$ correction to the Bohr-Sommerfeld quantization condition
and it also provides a natural interpretation of the limiting spectrum
 \eqref{limitspectrum}

The comparison with \eqref{ediff} also permits us to deduce an expression
for $m(\beta)$, the mass of the sinh-Gordon particle:
\begin{equation}\label{mass}
m({\beta})=\frac{8}{ \pi B}\sin \frac{\pi B}{ 4}.
\end{equation}
This is independently interesting. Previously, the same
expression for the mass has been
deduced via analytic continuation using a knowledge of the sine-Gordon
breather spectrum on the whole-line. However,  here it arrives naturally
within the context of the sinh-Gordon model itself. It appears
that once the model is defined in a restricted region by boundary
conditions which permit the existence of boundary states,  boundary
effects allow bulk parameters to be determined. Notice that periodic
boundary conditions, which are in some respects the most natural to
impose, and are certainly the traditional choice,
do not share this property.

\section{Discussion}
\label{s:discuss}

In this section we need to take another look at the
two descriptions of the boundary bound state spectrum.
Using what we have learned, the boundary states are described
by two different sets of angles which are linear functions of
$B$. From the WKB calculation we have the set $\rho_n$ given by
\begin{equation}\label{rhon}
\rho_n=\pi(1-a) +\frac{\pi}{2}\left(n+a-\frac{1}{2}\right)B,
\qquad n=0,1,2,3,\ \dots .
\end{equation}
The ground state corresponds to $\rho_0$ and lies in the spectrum
for all values of $B$. This is clear because as $B$ traverses its
range from 0 to 2, $\rho_0$ increases from $\pi(1-a)$ to $\pi/2$.
On the other hand, $\rho_n,\ n\ge 1$ corresponds to an excited state
which will leave the spectrum at some critical value of $B$ when
$\rho_n$ attains $\pi/2$.
Specifically, the critical couplings are given by,
\begin{equation}\label{Bcrit}
B_n^c= \frac{2(2a-1)}{2n+2a-1},\quad\hbox{or}\quad
\frac{\beta_n^{c\, 2}}{4\pi}=\frac{2a-1}{2n}.
\end{equation}

 The other description is derived from the bootstrap. Taking the
conjectured form of $E$, \eqref{E} with $N_0=1/2$, leads to another
set of angles $\psi_n$ defined by
\begin{equation}\label{psin}
\psi_n=\frac{\pi}{2}\left(2a-1 -(a+n-1)B\right),\quad n=1,2,3,\ \dots .
\end{equation}
Although these describe the same set of states via the bootstrap,
the angles are clearly very different. One striking difference
concerns the critical value of the coupling $B_n^{c\prime}$ at
which the state exits the spectrum. The angles \eqref{psin}
clearly decrease with increasing $B$ and the critical point is
reached when an angle vanishes. Thus, we have
\begin{equation}\label{Bcritp}
B_n^{c\prime}=\frac{2a-1}{a+n-1}, \quad\hbox{\rm or}\quad
\frac{\beta_n^{c\prime\, 2}}{4\pi}=\frac{2a-1}{2n-1}.
\end{equation}
The two critical points \eqref{Bcrit} and \eqref{Bcritp}
are similar but not  the same. Curiously, in terms of the
inverse coupling the difference is independent of $n$:
\begin{equation}
\frac{4\pi}{\beta_n^{c\, 2}}-
\frac{4\pi}{\beta_n^{c\prime\, 2}}=\frac{1}{2a-1}.\nonumber
\end{equation}
The fact that the two critical points are different needs
explanation. Unfortunately, we do not have a complete dynamical
explanation of this.
The problem is that a bound state appears to leave the spectrum
before the pole marking it in a reflection factor moves out
of range.

Consider the bound state with label $n$. At the associated
rapidity $i\psi_n$ there is a pole
in the two reflection factors $K_{n-1}$ and $K_n$. In the first
of these, the pole indicates the possibility of exciting the
state $n-1$ to the state $n$; in the second it indicates the
possibility of dropping from state $n$ down to state $n-1$. In
both cases, of course, the process is virtual, but in the second
the process corresponds to a `crossed' diagram.
Of the various parts in
\eqref{Kn}, the one which produces the cross-channel pole
is $(1-E+(n-1)B)$, one
of the factors in the denominator. At the critical coupling this
is cancelled by the factor $(1-E+nB)$ because, at the critical value
$B_n^c$, $2E-(2n-1)B=2$. This is consistent with a zero in the
S-matrix \eqref{s} at $i\pi B_n^c/2$ which contributes to the
cross-channel diagram. For values of the coupling between the
two critical values, the pole at $i\psi_n$ in $K_{n-1}$ needs
explanation.

That the pole indicating a bound state can persist beyond the value of the
coupling at which the bound state ceases to exist is a phenomenon familiar in
the breather spectrum of the bulk sine-Gordon model. In the notation we have
been using, we simply make the change $\beta\rightarrow i\beta$ and redefine
$B(i\beta)=-b$. Then, the $n$th breather leaves the spectrum at $b=2/n$. This
is typically signalled by the appearance of double poles in S-matrices, rather
than the pole position moving across the boundary of the physical strip.
However, in this case, the explanation for the pole beyond the critical
coupling lies in a Coleman-Thun mechanism using solitons.\footnote{We thank
Patrick Dorey for pointing this out to us.}

A second point we wish to discuss is the following. Given the
expression for $E$, \eqref{E} with $N_0=1/2$,
 we see immediately that if the
paramter $a$ is held fixed as  $B\rightarrow 2$, then $E\rightarrow 0$,
and every reflection factor \eqref{Kn} has the same limit
\begin{equation}\label{Klimit}
K_n\rightarrow -\frac{1}{(1)^2}.
\end{equation}
The latter is the classical reflection factor corresponding to
the boundary condition \eqref{bc} with $\ep_0=\ep_1=1$
\cite{Cor95}. It is natural to suppose that the same expression
for $E(\ep, \beta)$ will be appropriate for $\ep>0$ although
we cannot prove it. However, if it is the case, then almost
all reflection factors will have the property \eqref{Klimit}.
The exception to this is the symmetrical Dirichlet condition
whose reflection factor has the property
$K_D(4\pi /\beta)=K_D(\beta)$; that is, $K_D$ is self-dual. Apart
from noting the phenomenon we can offer no explanation as to why
one particular non-linear boundary condition should be singled
out to be  the limit point of almost all the
others. It will be interesting to discover if this remains so
after the complete analysis of the general case $\ep_0\ne\ep_1$.
It is perhaps worth remarking that this special boundary condition
is one of the two singled out in the supersymmetric version of
the model \cite{Inam95}
(for the other $\ep_0=\ep_1=-1$).  If the expression for $E$ is also
correct for $\ep>0$ then $E(1,\beta)\equiv 0$, indicating that
this specially symmetrical boundary condition also has a self-dual
reflection factor. Perhaps this is also true for the model with
supersymmetry.

%The sinh-Gordon model is the simplest example of an
%affine Toda field theory and a clear next step will be to
%try to extend the ideas of this article to other models
%with continuous boundary parameters. (These occur
%only among the theories based on the non-simply laced
%algebras).
%For models in the $ade$ series, the boundary parameters
%are discrete; for some choices of boundary parameters
%within the $a_n, \, n>1$ models there are conjectured
%reflection factors \cite{Cor94a, Del99}.

In this article, we have obtained the expression \eqref{E} for $E$ in terms of
the parameter $a$. However, there is an indication from work on higher
$a_n^{(1)}$ Toda theories that the renormalised boundary parameter is not $E$
itself but $G=E+B/2$. In these theories, for $n>1$, there is only a discrete
set of integrable boundary conditions and for many of these the reflection
factors are known \cite{Del99}. These reflection factors can be specialised to
the case $n=1$ which corresponds to the sinh-Gordon model and one obtains the
reflection factor \eqref{brsymm} at fixed (coupling constant independent)
values of $G=E+B/2$ rather than fixed values of $E$. Further motivation for
regarding $G=E+B/2$ as the physical boundary parameter comes from the study of
solitons in the sine-Gordon model on the half line \cite{sine}.

Finally, it must be said that the WKB method gives an all orders result in
terms of $\beta$ and is exact for the bulk sine-Gordon model. Again, we would
probably be surprised if that were not the case in the present setting. 
\vskip 1cm 

\noindent{\bf Acknowledgements} \vskip .5cm \noindent One of us (GWD) is
supported by an EPSRC Advanced Fellowship and both of us have been partially
supported by a TMR Network grant of the European Commission contract number
FMRX-CT-960012. EC thanks Network partners at ENS-Lyon and LAPP-Annecy, and the
Institute for Theoretical and Experimental Physics, Moscow, for providing
inspiring surroundings at various stages of this work. 
We are grateful to Patrick Dorey,
Brett Gibson and Gerard Watts for conversations.

\end{document}